# A Discrete Two-Dimensional Model of a Loaded Cantilever Influenced by Time-Dependent Forces

Gennady P. Berman[1], Vyacheslav N. Gorshkov[1,2], Vasily V. Kuzmenko[2], Umar Mohideen[3]

[1]Los Alamos National Laboratory, MS 213, Los Alamos, New Mexico 87545

[2]National Technical University of Ukraine "KPI," 37 Peremogy Ave., Bldg. 7, Kiev-56, 03056, Ukraine

[3]Department of Physics and Astronomy, University of California, Riverside, CA 92521

## Abstract

We developed a discrete two-dimensional model of a cantilever which incorporates the effects of inhomogeneity, the geometry of an attached particle, and the influence of external time-dependent forces. We provide a comparison between the solutions for our discrete model and its continuous limit. The rotational-vibrational mode is studied in detail. The results of numerical simulations demonstrate usefulness of our model for many applications when a cantilever has a complicated geometry and is affected by time-depended and distributed external forces.

## 1. Introduction

Micro-cantilevers represent the important parts of micro-electromechanical systems (MEMS), micro-devices and sensors [1-3] (see also references therein). In particular, the cantilevers are widely used in atomic force microscopy [4], in magnetic resonance force microscopy [5], for detection of small forces near surfaces [6,7,8], and in many other applications [9,10]. Very often, when used for a particular application, a cantilever must have a complicated geometry with inhomogeneously distributed masses and spring constants, been loaded with a semiconductor or a metallic particle in the region of its free end, and may experience a time-dependent force at its fixed end or along its length [11]. In all these situations the eigenfrequencies of the cantilever modes, the corresponding eigenfunctions, and the dynamical regimes of the cantilever response could be rather complicated and require adequate modeling and simulations. In simple cases, a cantilever can be modeled as a continuum medium by using a linear partial differential equation which describes the dynamics of the cantilever displacement, $z(x,t)$, where $x$ and $t$ are the spatial coordinate of the cantilever and time, respectively. In this case, the dynamics is actually linear and one-dimensional, and the coordinate, $x$, plays the role of an independent variable. At the same time, in many experimental situations, the dynamics of the cantilever can be two- and even three-dimensional and non-linear. Also, as mentioned above, the external time-dependent forces can be applied at a particular region of the cantilever. For all these purposes, a more general cantilever model should be developed which takes into account many relevant effects.

In this paper, we develop a discrete model of the cantilever. Our model takes into account (i) two-dimensional nonlinear cantilever oscillations, (ii) different time-dependent external forces acting on the cantilever such as forced oscillations of the fixed point, light pressure, and others, (iii) different distributions of masses and elastic constants which allow one to model different sorts of inhomogeneities, and, (iv) attached particles of different geometry. We describe our model in detail, and perform numerical simulations for different parameters and conditions. In particular, we demonstrate the numerical regimes for dynamics of the loaded cantilever under the influence of the external time-dependent force applied at the fixed end of the cantilever. The rotational-vibrational mode is analyzed in detail. We also compare the solutions for our discrete model with a continuous one, and discuss how to calibrate our model.

## 2. Eigenfrequencies of a loaded cantilever

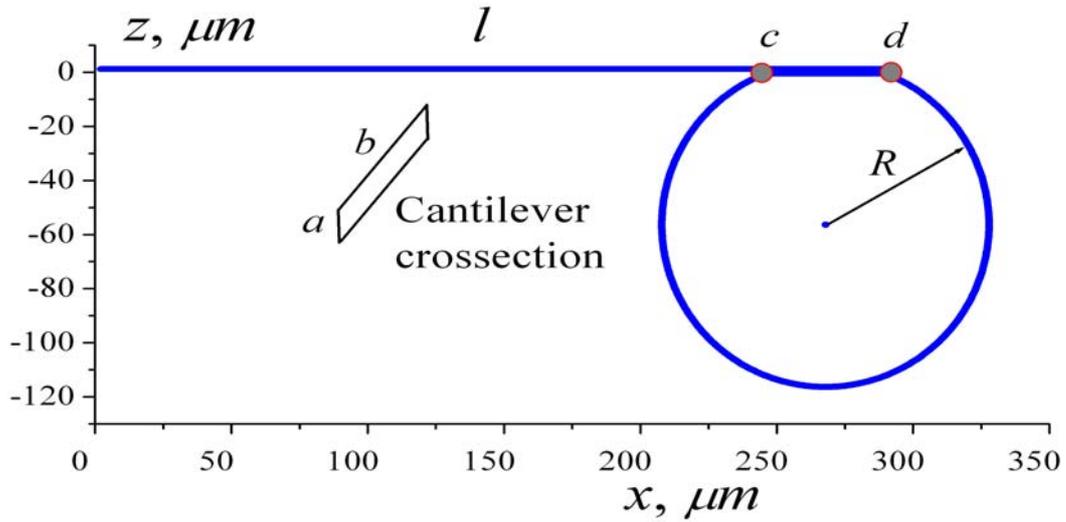

Fig. 1. Setup of a loaded cantilever.

**The statement of the problem.**
At the free end of a homogeneous cantilever of a length, $l$, cross-section, $S = ab$, and a mass, $m_0$, a sphere with a radius, $R$, and a mass, $m_{sp}$, is attached. Then, the problem is to find:
- The eigenfrequencies of this loaded cantilever, and
- its dynamics in the case when its fixed point ($x = 0$) experiences forced harmonic oscillations with the frequency, $\omega_{ex}$.

First, suppose that $S(x)$ and $I(x) = \frac{1}{12}a^3(x)b$ (the parameter which characterizes the elastic properties of the cantilever) are constant over its length, and the density, $\rho(x)$, is a step-function:

$$\rho = \rho_0 \text{ for } x < c, \text{ and } \rho = \rho_0 + \frac{m_{sp}}{l_f S} \text{ for } c < x < l,$$

where $l_f$ is the length of the segment, $cd$.

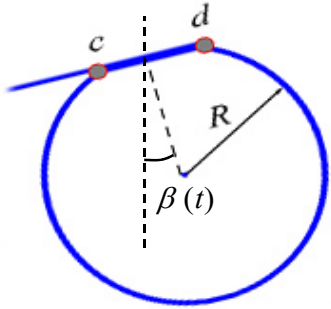

Fig. 2. Rotation of the sphere and shift of the center of gravity along the x-axis due to oscillations of the cantilever.

This is one of the possible approaches to describe the loaded cantilever. A concrete model for introducing the functions, $\rho(x)$ and $S(x)$, could be chosen by different methods. We will not discuss here these different approaches (which give similar results within a 2% accuracy), but present below the most natural approach.

In this case, the eigenfrequencies can be obtained from the equation for a cantilever written in an approximation of a continuous medium

$$\rho(x) S \frac{\partial^2 z}{\partial t^2} = EI \frac{\partial^4 z}{\partial x^4}, \ 0 \leq x \leq l, \tag{1}$$

where E is the Young's modulus. The boundary conditions are

$$z(0) = \left(\frac{\partial z}{\partial x}\right)_0 = \left(\frac{\partial^2 z}{\partial x^2}\right)_l = \left(\frac{\partial^3 z}{\partial x^3}\right)_l = 0. \tag{2}$$

Here, we assumed that the elastic properties of the cantilever do not change along its length, $E(x) = const$.

The system of equations (1) and (2) is valid for small oscillations, and does not take into account possible displacements of the cantilever along the *x*-axis. However, for large enough radius, $R$, of a sphere and its mass, $m_{sp}$, relative to the values, $l$ and $m_0$, the system of equations (1) and (2) will not adequately describe the real dynamics of the cantilever. Under the bending of the cantilever and its oscillations, the region of attachment of the sphere (a segment *cd* in Figs. 1 and 2) moves, and the center of gravity of the sphere shifts along the *x*-axis by the value, $\Delta x \approx R\beta(t)$. These additional rotational-vibrational motions (which was not taken into account in deriving of Eq. 1) could significantly modify the dynamics of the cantilever.

Below we develop a discrete model of the system "cantilever with a sphere" (CS) which allows us to take into account the complicated dynamics of the CS and also the nonlinear effects (if needed) at relatively large amplitudes of oscillations. The continuous model (1) and (2) will be used for testing and calibration of the discrete model and also for comparison of different solutions.

### 3. Discrete mathematical model of the cantilever

**A. Comparison of the results for continuous and discrete models for a homogeneous cantilever.** To simplify the description of our approach, first consider the case of a homogeneous cantilever. Then, to justify a discrete model, we compare its results for the eigenfrequencies with those of a continuous model. The eigenfrequencies are independent of the width of the cantilever, $b$. This follows from Eq. (1) which, after a substitution of $S = ab$ and $I = \dfrac{a^3 b}{12}$, takes the form

$$\rho_0 \frac{\partial^2 z}{\partial t^2} = E \frac{a^2}{12} \frac{\partial^4 z}{\partial x^4}, \quad 0 \le x \le l. \tag{3}$$

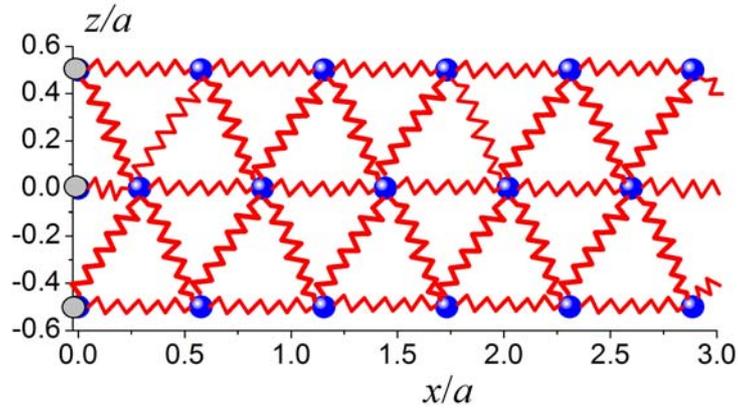

Fig. 3. A fragment of the cantilever (close to its fixed end) represented by a set of local oscillators (material points-blue spheres) and springs (red) which connect these oscillators. Gray circuits represent the points which either do not move or move according to a prescribed rule.

Introduce into Eq. (3) dimensionless variables, $\hat{x} = x/l$, $\hat{t} = t \times \dfrac{a}{l^2} \left( \dfrac{E}{12\rho_0} \right)^{1/2}$. Then, we have from Eq. (3)

$$\frac{\partial^2 z}{\partial \hat{t}^2} = \frac{\partial^4 z}{\partial \hat{x}^4}, \quad 0 \le \hat{x} \le 1. \tag{4}$$

Eq. (4) has a standard set of dimensionless eigenfrequencies $\{\bar{\omega}_n\}$, and corresponding eigenfunctions. The dimensional eigenfrequencies are

$$\omega_n = \bar{\omega}_n \times \frac{a}{l^2}\left(\frac{E}{12\rho_0}\right)^{1/2}. \qquad (5)$$

They depend on parameters, $a, l, E, \rho_0$. However, the dimensionless ratios, $\Omega_n = \omega_n/\omega_0$, do not depend on these parameters ($\omega_0$ is a fundamental mode of the homogeneous cantilever). Due to this fact, in what follows (when considering a loaded cantilever) we will also deal with the frequencies, $\{\Omega_n\}$, without connecting them to concrete values $a, l, E, \rho_0$.

In a discrete model, the cantilever is represented on the $xz$-plane by a system of material points and springs (see Fig. 3). In equilibrium, the distance between neighboring points is equal, and the cantilever is oriented along the $x$-axis. This discrete model takes into account the horizontal displacements of the separate elements of the cantilever and the nonlinear effects which appear at large enough amplitudes of oscillations. This model also allows us to analyze the action of the external time-dependent forces applied, in particular, at the fixed end of the cantilever.

One of the possibilities to find the eigenfrequencies is to initially bend the cantilever and allow it to oscillate. We describe below the dynamics of the cantilever which corresponds to a search for a self consistent solution of the system of second order differential equations (equations of motion) for coordinates of material points, $x_i$, $z_i$, in the field of elastic forces of the deformed springs. In order to fulfill the requirement for conservation of total energy, we used a discrete implicit scheme of second order of accuracy, by applying the iteration method for finding the solutions of these equations. To find the eigenfrequencies one can use the time-dependent coordinate of an arbitrary point, $z_i(t)$, in the time-interval, $t_{calc} \gg 2\pi/\omega_0$, expand it in the Fourier series with the coefficients, $A(\omega) \sim \int_0^{t_{calc}} z_i(t)e^{i\omega t}dt$, and build the function, $|A(\omega)|$.

The initial displacement of the cantilever was provided in the following way. Its fixed end was shifted along the $z$-axis by a value, $\Delta z \ll l$, during a small time-interval, $\tau$ ($\tau \ll 2\pi/\omega_0$). During this time, the free end of the cantilever does not move due to the finite inertia. At $t = \tau$ the cantilever is bent in accordance with the dynamics of the material points calculated for $0 < t < \tau$. For $t > \tau$ the cantilever develops free oscillations. The frequency spectrum in the discrete model depends on the mass of a single material point, $m^{(0)}$, the number of material points, $N$, the length, $l$, and the elastic coefficient of the springs, $k^{(0)}$. In fact, all these parameters define some effective values of the width, length, the Young's modulus, and the density of the cantilever, $a$, $l$, $E$, $\rho_0$ ($l \gg a$). Independent of the choice of these parameters, the eigen frequencies, $\{\Omega_n\}$, will remain the same.

To verify the correctness of the discrete model in the simple case of a homogeneous cantilever, we have chosen the parameters, $a = 1\mu m$, $l = 350\mu m$. Then, in each outermost horizontal row we took 607 material points, and in the middle row – 608. The mass of each point and the elastic coefficient of each spring were taken as an unity. The fixed end of the cantilever was shifted during a small time-interval by $\Delta z = 5\mu m$, and then was fixed. Some pictures of the oscillating cantilever are presented in Fig. 4. The eigenfrequencies derived from the discrete model, $\{\Omega_n\}$, are in good correspondence with the eigen frequencies derived from Eq. (4) for a continuous model. These eigenfrequencies were obtained from the Fourier spectrum for the coordinate, $z(t)$, of the most right point in the middle row.

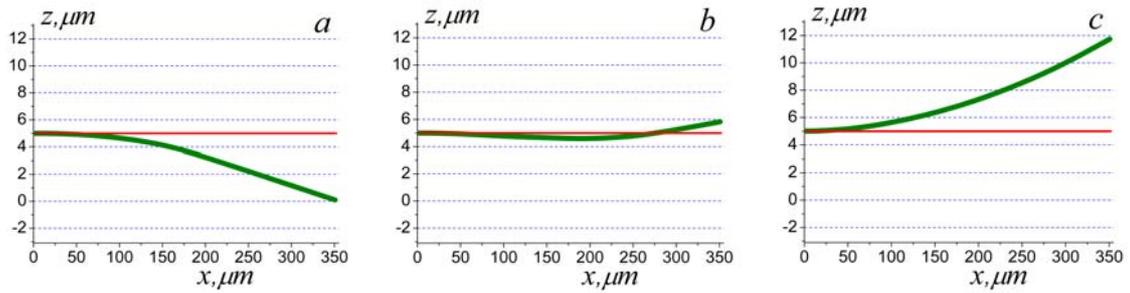

Fig. 4. The shape of the cantilever at different moments in time; ($a$) the moment of the maximal displacement of the free end of the cantilever, $\Delta z = -5\mu m$; ($b$) the shape of the cantilever at an intermediate moment of time, and ($c$) the maximal displacement with $\Delta z = 5\mu m$.

**Note.** The eigenfrequencies from Eq. (4) are $\bar{\omega}_n = \lambda_n^2$, where $\lambda_n$ are the solutions of the equation: $\cos(\lambda_m) ch(\lambda_m) + 1 = 0$.
$\bar{\omega}_0 = 3.516$; $\bar{\omega}_n = 22.03$; $61.69$; $120.9$; $199.85$; $298.55$, $n = 1, 2, 3, 4, 5$. Correspondingly, for the first five harmonics we have

$$\{\Omega_n\} = 6.26;\ 17.54;\ 34.38;\ 56.84;\ 84.9;\ n = 1, 2, 3, 4, 5. \tag{6}$$

**B**. **Simplified models of a loaded cantilever.** A comparison of the results for a loaded cantilever following from continuous and discrete models will allow us to test and calibrate the discrete model.

Consider a cantilever with a constant cross section along its length, and the mass of the attached particle is homogeneously distributed in the segment, *cd* (simplified model of a loaded cantilever).

In the continuous model this corresponds to a stepwise increase in the cantilever density. Its dimensionless value, $\hat{\rho}(\hat{x}) = \rho(\hat{x})/\rho_0$, is equal to 1 outside the region *cd* (See Fig.1) and $\hat{\rho}(\hat{x}) = 1 + m_{sp}/(\rho_0 S l_f)$ for $1 - l_f/l < x < 1$ ($l_f$ is the length of the segment *cd* ). Eq. (4) must be substituted by the equation

$$\frac{\partial^2 z}{\partial \hat{t}^2} = \frac{1}{\hat{\rho}(\hat{x})} \frac{\partial^4 z}{\partial \hat{x}^4}, \quad \hat{\rho}(\hat{x}) = \begin{cases} 1, \ 0 < \hat{x} < 1 - \hat{l}_f; \ \hat{l}_f = l_f/l, \\ 1 + \dfrac{m_{sp}}{m_0 \hat{l}_f}, \ 1 - \hat{l}_f < \hat{x} < 1. \end{cases} \quad (7)$$

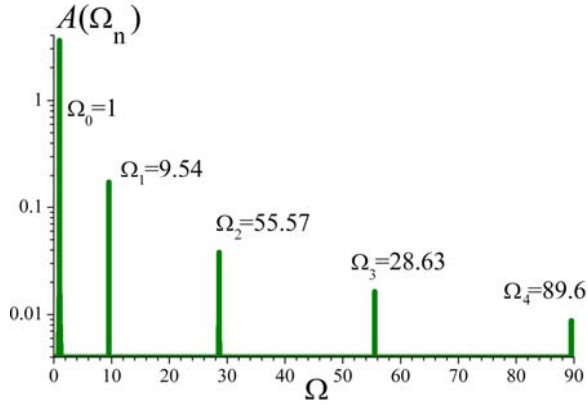

Fig. 5. The eigenfrequencies, $\{\Omega_n\}$, of a loaded cantilever calculated for a simplified discrete model with $\hat{l}_f = 0.05$, $m_{sp}/m_0 = 0.72$. Initial excitation of the cantilever was provided by a short displacement of the fixed end by a value, $\Delta z = l/100$.

The eigenfrequencies for Eq. (7) mainly depend on the ratio, $m_{sp}/m_0$, if $\hat{l}_f \ll 1$. The corresponding numerical procedure of calculation of the eigenfrequencies is presented in the Appendix.

In the discrete model, the mass of material particles located in the segment, $1 - \hat{l}_f < \hat{x} < 1$, should be increased by the factor, $\left(1 + \dfrac{m_{sp}}{m_0} \dfrac{N}{N_f}\right)$, where $N$ is a total number of points, and $N_f$ is their number in the region, $1 - \hat{l}_f < \hat{x} < 1$. In Fig. 5, the eigenfrequencies, $\{\Omega_n\}$, are presented for the discrete model with $\hat{l}_f = 0.05$, $m_{sp}/m_0 = 0.72$.
In the continuous model, $\Omega_n = 9.42;\ 28.52;\ 56.26;\ 91$ for $n = 1, 2, 3, 4$, which is in a good agreement with the results obtained for the discrete model.

The fundamental frequency of a loaded cantilever is smaller than for unloaded one. For the chosen parameters (see Fig.5) the fundamental frequency is reduced by the factor of 1.9 independent of the scheme used in the numerical simulations.

After testing of the discrete model, we will describe below the numerical simulation of the eigenfrequencies of a loaded cantilever with an attached sphere.

**C. Simulation of the frequency spectrum in the discrete model with an attached sphere.** In Fig. 6, the configuration of the right end of the cantilever is shown. The motion of the material particles which do not belong to the region, $cd$, is determined by the deformation of the connected springs. The points belonging to the region, $cd$, have rigid connections to the sphere. The motion of these points depends on the dynamics of the whole sphere. A set of the springs connected to the segment, $cd$, create a resulting elastic force, $\vec{f}$, and the rotation moment, $\vec{M}$, relatively to the center of the sphere. The motion of the center of gravity of the sphere, $\vec{r}_{0,sp}(t)$, and the angle of its rotation, $\beta(t)$, are described by the equations

$$m_{sp}\frac{d^2\vec{r}_{0,sp}}{dt^2} = \vec{f}, \quad I_{sp}\frac{d^2\beta(t)}{dt^2} = M, \qquad (8)$$

where, $I_{sp}$ is the moment of inertia of the attached particle relative to its center of mass. One can use an approximate relation, $I_{sp} = 0.4 m_{sp} R^2$, neglecting a small part of the sphere.

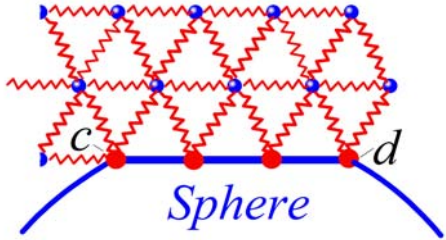

Fig. 6. Presentation of the right end of the loaded cantilever in a discrete model. Red points belonging to the segment $cd$, have a rigid connection with the sphere.

Eqs. (8) and equations of motion for material points (which do not belong to the region $cd$) should be solved self-consistently. At each step of the numerical simulation of the whole system of equations, the coordinates of the connected points (belonging to a segment $cd$) are calculated by using the coordinates of $\vec{r}_{0,sp}$ and the angle of rotation, $\beta$.

The solution of the problem depends on the ratios, $m_{sp}/m_0$, $R/l$, and the parameter, $\hat{l}_f$. Below we discuss the numerical results for the cantilever with typical parameters used in experiments:

$$a = 1\mu m, \; l = 290\mu m, \; R = 60\mu m, \; m_{sp}/m_0 = 0.75,$$

and $\hat{l}_f \approx 0.15$ (in the region, $cd$, the number of material points, $N_f = 77$).

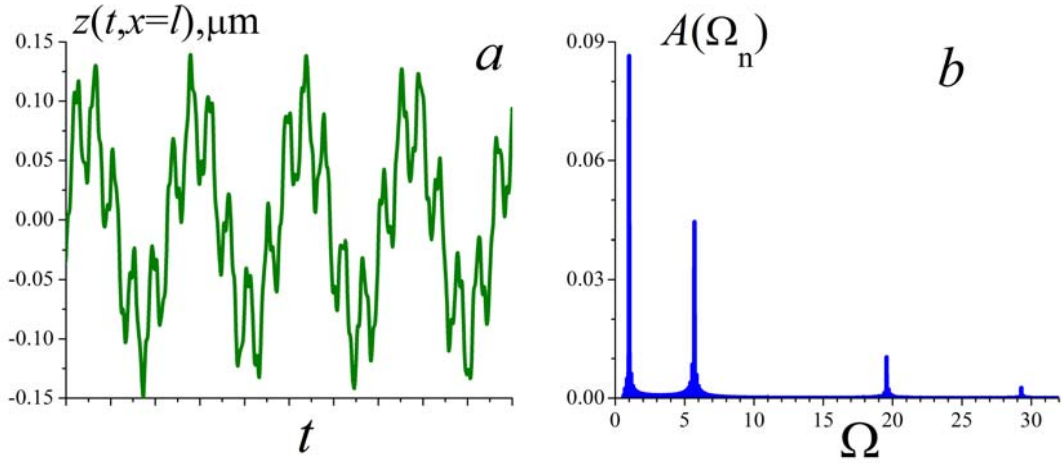

Fig.7. (*a*): A fragment of the trajectory of the end of the cantilever, $z(t, x = l)$. One can see that the oscillations at the fundamental frequency are modulated by the rotational oscillations of the sphere at frequency, $\Omega_{rot} \approx 5.8$. (*b*): The frequency spectrum of $z(t, x = l)$, where $z(t, x = l)$ is the averaged value for the three right most material points (see Fig. 2).

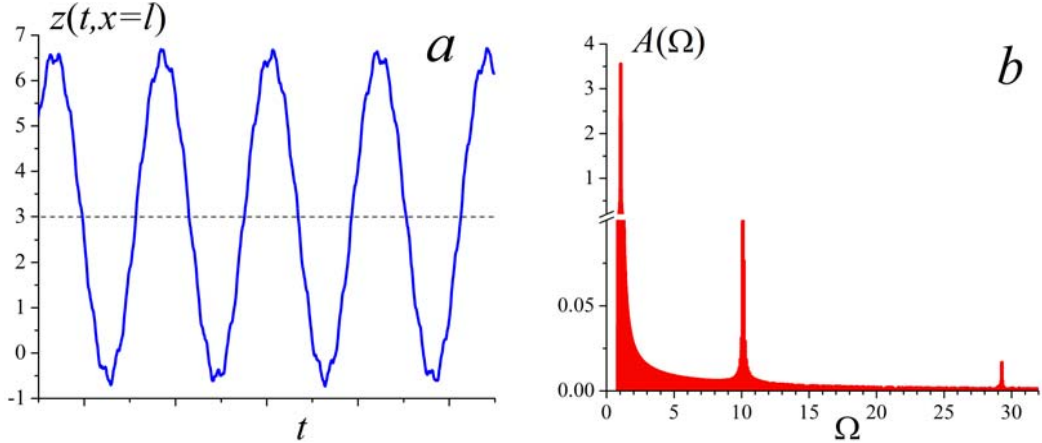

Fig. 8. Results of simulations using a simplified model (sub-section B); (*a*): A fragment of the trajectory of the cantilever, $z(t, x = l)$. (*b*) A frequency spectrum of $z(t, x = l)$.

Analyzing the configuration shown in Fig. 6 one can assume that in the spectrum of the cantilever new modes will appear which are related to the oscillations of the sphere relative to its region of connection, *cd* (right-left oscillations in Fig. 6). Under this dynamics, the end of the cantilever experiences small oscillations along the *z*-axis, and its central part bends significantly, shifting from the equilibrium state. To demonstrate these modes, we excite the cantilever by a short-time displacement of its left end along the *x*- axis, $\Delta x \approx 0.37 \mu m$ (previously we excited the cantilever by a displacement along the $z$-axis). In the present case, the initial shift in the $x$-direction is accompanied by bending of the cantilever (due to the presence of the sphere, whose center of gravity is shifted along the $x$-axis) and by the excitation of oscillations along the $z$-direction. (See Fig. 7.)

For comparison, we present the results of a simulation (Fig. 8) for a simplified model (see sub-section B). During some initial time-interval the cantilever was excited by oscillating its left end along the $z$-axis. Then, the cantilever produced free oscillations.

Note, that the absolute values of the fundamental frequencies in these two models are practically the same (in the "total" model the frequency, $\bar{\omega}_0$, is only 3% larger than for the simplified model).

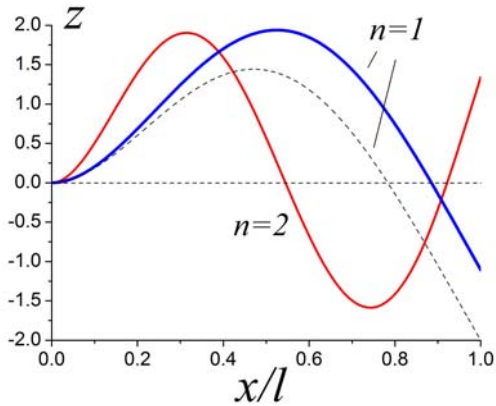

Fig. 9. The eigenfunctions of the loaded cantilever calculated for a simplified continuous model. Dotted curve corresponds to the calculations for the unloaded cantilever.

Let us analyze the data presented in Figs 7b and 8b. In both spectrums the frequency, $\Omega \approx 29.5$, is present, which corresponds to the second harmonics (seen also in Fig. 5). Significant modifications are observed in the frequency spectrum in Fig. 7b. Between the fundamental mode and the "second" mode two eigenfrequencies are present in contrast to the data shown in Fig. 8b. The physical explanation of this is the following.

The new mode, $\Omega_{rot} \approx 5.8$, corresponds to the rotational-vibrational motion of the sphere relative to its region of attachment. It is natural to call the frequency, $\Omega \approx 19.5$, as before, the "first" higher harmonic as the corresponding oscillations possess a similar physical nature as the oscillation on the first higher harmonic, $\Omega_1 \approx 10$, in the spectrums presented in Fig. 8*b* and Fig. 5. We analyze now the physical mechanism which leads to an increase of the frequency, $\Omega_1$, in Fig. 7.

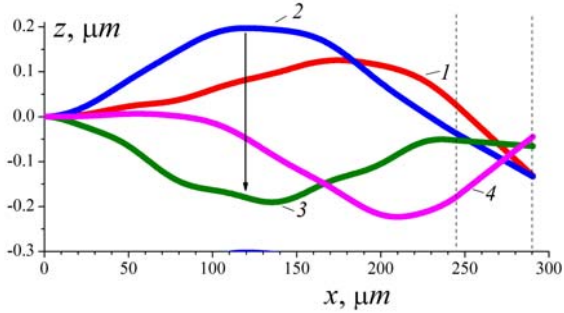

Fig. 10. The shape of the cantilever in different moments of time. 1: the moment of the end of the horizontal shift along the *x*-axis, $t = 5\Delta t$. 2: the moment of the maximal bending of the cantilever "up", $t = 8\Delta t$. 3,4: $t = 15\Delta t$ and $t = 22\Delta t$. $\Delta t = T/240$, $T$ is the period of oscillations at the fundamental frequency.

Consider the eigenfunctions of the loaded cantilever in the simplified model presented in Fig. 9. One can see that the oscillations of the end of the cantilever are also accompanied by rotations (the angle, $\beta(t)$, relative to the *z*-axis changes in time). The loaded part of the cantilever (in a simplified model) has relatively small inertial moment when the loaded particle is localized in a small region. So, in this case, the dynamics of the slope of the loaded part is synchronized with the oscillations of the middle part of the cantilever. When the middle part bends up, the slope of the end is negative, and vice versa. The same is true for mode 2.

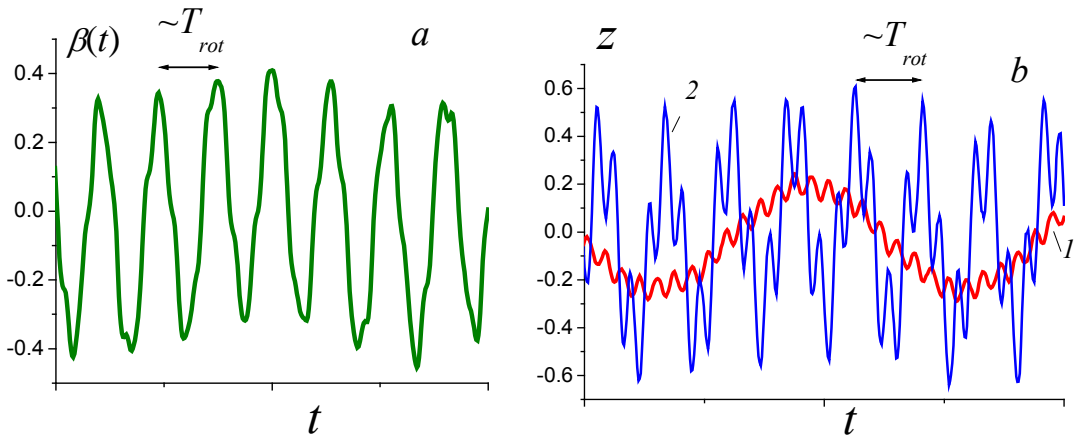

Fig. 11. The dynamics of the cantilever with an "empty" sphere (with the same mass). (*a*): The angle, $\beta(t)$, of rotation (in degrees) relative to the point of attachment, see Fig. 2. (*b*): 1 (red): oscillations of the end of the cantilever; 2 (blue): oscillations of its central part. The time-interval corresponds approximately to 1.5 periods of oscillations on the fundamental mode. The length of attachment of the sphere to the cantilever is reduced from 77 points to 55.

Now consider the dynamics for the total discrete model (Fig. 10). Fast variations of the bending from "up" to "down" (curves 2 and 3) are not followed by the end of the cantilever with a large moment of inertia. The slope of the end relative to the *x*-axis remains negative (curves 2 and 3). Then, the effective length of the cantilever, $l_{eff}$, that undergoes the oscillations (which are physically analogous to the first harmonic) is reduced. Consequently, the frequency of these oscillations increases. As one can see from Fig. 7b, it increased by a factor two: $\Omega_1 \sim 19.5$.

For the second mode of oscillations (see Fig. 9, $n = 2$) the eigenfunction has a node at the central part of the cantilever. In this case, the value of the moment of inertia of the sphere at the end of the cantilever has a small influence on the frequency of oscillations (see Fig. 7b and 8b- $\Omega_2 \sim 29.5$).

Note, that the amplitudes of the modes can change significantly depending on the moment of inertia of the sphere and the region of measuring of the cantilever oscillations, $z(t)$, along the $x$-axis.

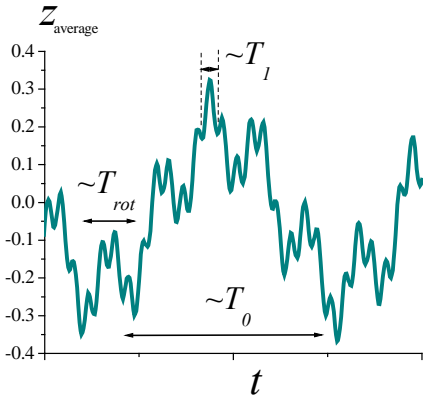

Fig. 12. The dynamics of $z(t)$ averaged over the attachment region of the sphere to the cantilever.

Above we assumed that the mass of the sphere is homogeneously distributed over its volume. If the sphere is mainly empty, and the mass is distributed only in a thin layer near its surface, the moment of inertia of the sphere relative to its center is, $I_{sp} = m_{sp} R^2$. This is by a factor 2.5 higher than the value we used above, $I_{sp} = 0.4 m_{sp} R^2$, in Eq. (8). In Fig. 11 the dynamics of the cantilever is shown after a short-pulse excitation by a horizontal displacement.

In Fig. 11a only a rotational-vibrational mode is visualized with the frequency, $\Omega_{rot} \approx 5$. However, when measuring the oscillations of the cantilever end, (Fig. 11b-red) one can see only the fundamental mode and a modified "first" mode. The visual estimate of this frequency gives, $\Omega_1 \approx 18$. A rotational-vibrational mode is not seen on this curve (opposite to the analogous curve in Fig. 7a, where the rotational-vibrational mode is clearly seen). At the central part of the cantilever, (Fig. 11b-blue) the deviations dominate and are related to the bending of the cantilever resulting from the oscillation of the sphere along the $x$-axis. These oscillations are modulated by high-frequency oscillations related to the "first" mode.

All three first modes are visually distinguished simultaneously if the value $z_{average}(t)$ is averaged over the region of attachment (see Fig. 12). An increase of the moment of inertia of the sphere resulted in a decrease of (i) the frequency, $\Omega_{rot}$, to a value of 4.8 (in Fig. 7b, $\Omega_{rot} = 5.8$) and (ii) by ~ 6% of the frequency of the fundamental mode. A reduction of the attachment region (increase of $l_{eff}$) leads to a decrease of $\Omega_1 = 17.8$ (19.5 in Fig. 7b). Finally, a modification of the moment of inertia of the sphere reveals itself mainly on the rotational-vibrational mode, which is physically reasonable from the above-described mechanism of such oscillations.

**Conclusion**

We have discussed a discrete model of the cantilever, which allows one to perform numerical simulations for a wide class of experimental conditions and parameters. With this model, one can easily take into account (i) two-dimensional cantilever oscillations, (ii) different time-dependent external forces acting on the cantilever such as forced oscillations of the fixed point, light pressure, etc, (iii) different distributions of masses and elastic constants which allow one to model different sorts of inhomogeneities, and (iv) attached particles of different geometry. Our results demonstrate that modeling of the cantilever by using three rows of material points (as in Figs 3 and 6) is a good approximation of the continuous medium. We also tested numerically the discrete model with five rows of material points. The results are practically identical to the model with the three rows. We believe that the discrete model presented in this paper will be useful for many applications in cantilever and MEMS technologies when specific modeling and simulations are required.

**Acknowledgement**

This work was carried out under the auspices of the National Nuclear Security Administration of the U.S. Department of Energy at Los Alamos National Laboratory under Contract No. DE-AC52-06NA25396. This project was funded by the UC Lab Fees Research Program. The EU-IndiaGRID2 project (European FP7 e-Infrastructure grant, contract no. 246698, www.euindiagrid.eu) is acknowledged for the use of its grid infrastructure.

**Appendix**

The equation

$$\hat{\rho}(\hat{x})\frac{\partial^2 z}{\partial \hat{t}^2} = \frac{\partial^4 z}{\partial \hat{x}^4}, \tag{a1}$$

has to be solved with the boundary conditions,

$$z(0) = \left(\frac{\partial z}{\partial \hat{x}}\right)_0 = \left(\frac{\partial^2 z}{\partial \hat{x}^2}\right)_{x=1} = \left(\frac{\partial^3 z}{\partial \hat{x}^3}\right)_{x=1} = 0. \tag{a2}$$

The eigenfunctions, $f_n(\hat{x})$, and the eigenfrequencies, $\bar{\omega}_n$, of the cantilever satisfy the equation

$$\hat{\rho}(\hat{x})\bar{\omega}_n^2 f_n(\hat{x}) = \frac{\partial^4 f_n(\hat{x})}{\partial \hat{x}^4}. \tag{a3}$$

The condition of orthogonality can easily be derived from (a3) [*]

$$\int_0^1 \hat{\rho}(\hat{x}) f_n(\hat{x}) f_m(\hat{x}) d\hat{x} = 0. \tag{a4}$$

We normalize the eigenfunctions, $f_n(\hat{x})$, in a standard way: $\dfrac{\int_0^1 \hat{\rho}(\hat{x}) f_n^2(\hat{x}) d\hat{x}}{\int_0^1 \hat{\rho}(\hat{x}) d\hat{x}} = 1$.

To find the eigenfunctions, $f_n(\hat{x})$, of the nonuniform cantilever we use the expansion over the eigenfunctions, $\varphi_k(\hat{x})$, of the uniform cantilever

$$f_n(\hat{x}) = \sum_{k=1}^M c_{kn} \varphi_k(\hat{x}), \quad \int_0^1 \varphi_i(\hat{x}) \varphi_j(\hat{x}) d\hat{x} = \delta_{ij}. \tag{a5}$$

For the eigenfunctions, $\varphi_k(\hat{x})$, of the uniform (unloaded) cantilever and its eigenfrequencies, $\omega_k$, the following conditions are satisfied:

$$\frac{\partial^4 \varphi_k(\hat{x})}{\partial \hat{x}^4} = \lambda_k^4 \varphi_k(\hat{x}), \quad \cos(\lambda_k) ch(\lambda_k) + 1 = 0, \quad \lambda_k^2 = \bar{\omega}_k.$$

For the fundamental mode of the unloaded cantilever we have

$$\bar{\omega}_0 = 3.516. \tag{a6}$$

We take the number of basis functions, $M = 50$, which provides an accurate approximation for the eigenfunctions of the nonuniform cantilever. In particular, we have verified the conditions of orthogonality (a4) for the normalized eigenfunctions, $f_n(\hat{x})$ - the values of the integral in the left side of Eq. (a4) don't exceed 0.02. In our computations for a given value of $n$ we have found both the eigenvectors, $c_{kn}$ ($1 \leq k \leq M$), and eigenvalues, $\bar{\omega}_n^2$, of the matrix, $\alpha_{km}$

---

[*] *By* using integration by parts, $\bar{\omega}_n^2 \int_0^1 \hat{\rho}(\hat{x}) f_n(\hat{x}) f_m(\hat{x}) d\hat{x} = \int_0^1 \dfrac{\partial^2 f_n(\hat{x})}{\partial \hat{x}^2} \dfrac{\partial^2 f_m(\hat{x})}{\partial \hat{x}^2} d\hat{x}$. Similarly, $\bar{\omega}_m^2 \int_0^1 \hat{\rho}(\hat{x}) f_m(\hat{x}) f_n(\hat{x}) d\hat{x} = \int_0^1 \dfrac{\partial^2 f_m(\hat{x})}{\partial \hat{x}^2} \dfrac{\partial^2 f_n(\hat{x})}{\partial \hat{x}^2} d\hat{x} = \bar{\omega}_n^2 \int_0^1 \hat{\rho}(\hat{x}) f_n(\hat{x}) f_m(\hat{x}) d\hat{x}$. If $n \neq m$, then $\bar{\omega}_n \neq \bar{\omega}_m$, and Eq. (a4) must hold.

$$\bar{\omega}_n^2 c_{kn} = \sum_{m=1}^{M} \alpha_{km} c_{mn}, \tag{a7}$$

$$\alpha_{km} = \int_0^1 \frac{\varphi_k(\hat{x})}{\hat{\rho}(\hat{x})} \frac{\partial^4 \varphi_m(\hat{x})}{\partial \hat{x}^4} d\hat{x} = \lambda_m^4 \int_0^1 \frac{\varphi_k(\hat{x})\varphi_m(\hat{x})}{\hat{\rho}(\hat{x})} d\hat{x}. \tag{a8}$$